\documentclass[aps,graphicx,eqsecnum,showpacs]{revtex4}
\begin{document}
\hspace*{14cm} {\large US-03-10}
\vspace{10mm}

\title{Universal Texture of Quark and Lepton  Mass Matrices \\ 
with an Extended Flavor $2\leftrightarrow 3$ Symmetry }

\author{Yoshio Koide}
\affiliation{
Department of Physics, 
University of Shizuoka, 52-1 Yada, Shizuoka, 422-8526 Japan \\
E-mail: koide@u-shizuoka-ken.ac.jp}

%\date{\today}

\title{Universal Texture of Quark and Lepton  Mass Matrices \\ 
with an Extended Flavor $2\leftrightarrow 3$ Symmetry }

\author{Yoshio Koide}
\affiliation{
Department of Physics, 
University of Shizuoka, 52-1 Yada, Shizuoka, 422-8526 Japan \\
E-mail: koide@u-shizuoka-ken.ac.jp}

\date{\today}

\begin{abstract}
Against the conventional picture that the mass matrix forms 
in the quark sectors will take somewhat different structures 
from those in the lepton sectors, on the basis of an idea that 
all the mass matrices of quarks and leptons have the same texture, 
a universal texture of quark and lepton mass matrices is proposed 
by assuming a discrete symmetry Z$_3$ and an extended flavor  
$2\leftrightarrow 3$ symmetry. The texture is described by three 
parameters (including phase parameter).  According to this ansatz, 
the neutrino masses and mixings are investigated.
\end{abstract}


\maketitle
%%%%%%%%%%%%%%%%%%%%%%%%%%%%%%%%%%%%%%%%%%%%%%%%%%%%%%%%%%%%%%%%%%%%%%
%\begin{multicols}{2}

\section{Introduction}

From the point of view of the quark and lepton unification,
an idea that their matrix forms are described by a universal
texture is very attractive.
Especially, against the conventional picture that 
the mass matrix forms in the
quark sectors will take somewhat different structures from those in the 
lepton sectors, it is interesting to investigate  whether or not
all the mass matrices of quarks and leptons can be described
in terms of the same mass matrix form as in the neutrinos.
Recently, a quark and lepton mass matrix model based on a discrete symmetry
Z$_3$ and a flavor $2\leftrightarrow 3$ symmetry has been proposed 
\cite{univtex}. In the model (we will refer it as Model I hereafter), the
quark and lepton mass matrices $M_f$ are given by the texture
\begin{equation}
M_f = P_f  \widehat{M}_f  P_f  \ ,
%\eqno(1.1)
\end{equation}
where $\widehat{M}_f$ is a real matrix with a form
\begin{equation}
\widehat{M}_f 
=
a_f \left(
\begin{array}{ccc}
0 & 1 & 1 \\
1 & 0 & 0 \\
1 & 0 & 0
\end{array} \right) 
+
b_f \left(
\begin{array}{ccc}
0 & 0 & 0 \\
0 & 1 & x_f \\
0 & x_f & 1
\end{array} \right)
= \left(
\begin{array}{ccc}
0 & a_f & a_f \\
a_f & b_f & b_f x_f \\
a_f & b_f x_f & b_f
\end{array} \right)   \ ,
%\eqno(1.2)
\end{equation}
and $P_f$ is a phase matrix defined by
\begin{equation}
P_f = {\rm diag} (e^{i \delta^{f}_1},\ e^{i \delta^{f}_2}, \
e^{i \delta^{f}_3}) \ .
%\eqno(1.3)
\end{equation}
(As seen in the expression in Eq.~(1.2), the model 
is essentially based on a two Higgs doublet model.)
As we see in the next section, 
Model I can give interesting
results in the quark and lepton mass matrix phenomenology.
However, in the model, the $2\leftrightarrow 3$ symmetry has
been required only for the mass matrix $\widehat{M}_f$, not
for the fields $\nu_{Li}$.  The phase matrix $P_f$ in Eq.~(1.1),
which breaks the $ 2\leftrightarrow 3$ symmetry, 
has been introduced from a phenomenological point of view.

In the present model, we propose a universal texture of quark
and lepton mass matrices
\begin{equation}
M_f = a_f \left(
\begin{array}{ccc}
0 & e^{-i\phi_f} & 1 \\
e^{-i\phi_f} & 0 & 0 \\
1 & 0 & 0
\end{array} \right) 
+
b_f \left(
\begin{array}{ccc}
0 & 0 & 0 \\
0 & e^{-2i\phi_f}1 & 1 \\
0 & 1 & 1
\end{array} \right)  \ ,
%\eqno(1.4)
\end{equation}
which is similar to the matrix (1.1), but does not
include the phenomenological phase matrix (1.3).
(The phases $e^{-i\phi}$ and $e^{-2i\phi}$ 
in the matrix (1.4) are introduced by an ``extended
$2\leftrightarrow 3$ symmetry which will be discussed
in Sec.~III.)
In comparison with Model I where the texture (1.1) has 
5 parameters, $a_f$, $b_f$, $x_f$, $\delta_1^f -\delta_2^f$ 
and $\delta_2^f -\delta_3^f$, the present model (1.4) has
only 3 parameters, $a_f$, $b_f$ and $\phi_f$, so that
the 3 mass eigenvalues can completely determine the
3 parameters $a_f$, $b_f$ and $\phi_f$.
As a result, for example, we will obtain a prediction
\begin{equation}
|V_{cb}| \simeq \frac{m_s}{m_b} + \frac{m_c}{m_t}  \ ,
%\eqno(1.5)
\end{equation}
differently from Model I, where $|V_{cb}|$ has been given
by $|V_{cb}|=\cos(\delta_3-\delta_2)/2$, where 
$\delta_i=\delta_i^u -\delta_i^d$, and the value
$|V_{cb}|$ has been freely adjustable by the parameter 
$\delta_3-\delta_2$.

In the next section, Sec.~II, we will give a brief review of
Model I, because the present model is closely related to
Model I.
In Sec.~III, by introducing an extended flavor 
$2\leftrightarrow 3$ symmetry, 
we will propose a new universal texture of quark and lepton mass 
matrices and we will investigate quark mass matrix phenomenology. 
In Sec.~IV, we will discuss the neutrino mass matrix $M_\nu$ 
on the basis of the new universal texture. 
Prediction of $\sin^2 2\theta_{atm}$ and $|(V_{\ell})_{13}|^2$ are 
given only in terms of the charged lepton mass ratios, independently of the
parameters in $M_\nu$. Predictions of 
$R=\Delta m^2_{solar} / \Delta m^2_{atm}$
and $\tan^2 \theta_{solar}$ depends on two adjustable parameters 
in $M_\nu$. 
We will give predictions for some typical values of the parameters. 
Finally, Sec.~V is devoted to a summary and discussion.

%%%%%%%%%%%%%%%%%%%%%%%%%%%%%%%%%%%%%%%%%%%%%%%%%
\section{Two Higgs doublet model with a Z$_3$ symmetry}

In the present section, we give a brief review of Model I \cite{univtex}.

We assume that under a discrete symmetry Z$_3$, 
the quark and lepton fields $\psi_L$,
which belong to $10_L$, $\overline{5}_L$ and $1_L$ of SU(5) 
($1_L={\nu}_R^c$),
are transformed as
\begin{equation}
\psi_{1L} \rightarrow \psi_{1L} , \ \
\psi_{2L} \rightarrow \omega \psi_{2L} , \ \
\psi_{3L} \rightarrow \omega \psi_{3L}, 
%(2.1)
\end{equation}
where $\omega=e^{2i\pi/3}$.
[Although we use a terminology of SU(5), at present,
we do not consider the SU(5) grand unification.]
Then, the bilinear terms $\overline{q}_{Li} u_{Rj}$, 
$\overline{q}_{Li} d_{Rj}$, $\overline{\ell}_{Li} \nu_{Rj}$, 
$\overline{\ell}_{Li} e_{Rj}$ and $\overline{\nu}_{Ri}^c \nu_{Rj}$
[$\nu_R^c =(\nu_R)^c =C \overline{\nu_R}^T$ and
$\overline{\nu}_R^c =\overline{(\nu_R^c)}$] 
are transformed as follows:
\begin{equation}
\left( 
\begin{array}{ccc}
1 & \omega^2 & \omega^2 \\
\omega^2 & \omega & \omega \\
\omega^2 & \omega & \omega \\
\end{array} \right) \ .
%(2.2)
\end{equation} 
Therefore, if we assume two SU(2) doublet Higgs scalars $H_A$ and $H_B$, 
which are transformed as
\begin{equation}
H_A \rightarrow \omega H_A , \ \ \ H_B \rightarrow \omega^2 H_B , 
%(2.3)
\end{equation}
we obtain the mass matrix form 
\begin{equation}
M_f = 
\left(
\begin{array}{ccc}
0 & a_{12}^f & a_{13}^f \\
a_{12}^f & 0 & 0 \\
a_{13}^f & 0 & 0
\end{array} \right)
\langle H^0_A \rangle + 
\left(
\begin{array}{ccc}
0 & 0 & 0 \\
0 & b_{22}^f & b_{23}^f \\
0 & b_{23}^f & b_{33}^f
\end{array} \right)
\langle H^0_B \rangle\ .
%\eqno(2.4)
\end{equation}
In addition to the Z$_3$ symmetry, we assume a 
$2\leftrightarrow 3$ symmetry for
the matrix $\widehat{M}_f$ which is given by Eq.~(1.2).
(The $2\leftrightarrow 3$ symmetry does not mean the
permutation $2\leftrightarrow 3$ symmetry for the fields
$\psi_{2L}$ and  $\psi_{3L}$.)
Hereafter, for simplicity, we will sometimes drop the index $f$ and 
denote $a^f \langle H^0_A \rangle$ and $b^f \langle H^0_B \rangle$ as
$a_f$ and $b_f$, respectively.
Then, we obtain the universal texture (1.1) with Eq.~(1.2) 
for the quark and lepton mass matrices.

Since the present model has two Higgs doublets horizontally, 
flavor-changing neutral currents (FCNCs) are, in general, caused by
the exchange of Higgs scalars.
However, this FCNC problem is a common subject 
to be overcome not only in the present model but also in most 
models with two Higgs doublets. 
The conventional mass matrix models 
based on a GUT scenario cannot give realistic mass 
matrices without assuming more than two Higgs scalars
\cite{howmanyH}.
Besides, if we admit that two such  scalars remain until
the low energy scale, the well-known beautiful coincidence
of the gauge coupling constants at $\mu \sim 10^{16}$ GeV
will be spoiled.
For these problems, 
we optimistically consider 
that only one component of the linear combinations 
among those Higgs scalars survives at the low energy scale 
$\mu=m_Z$, while the other component is decoupled at
$\mu < M_X$.  
Such an optimistic scenario
in a multi-Higgs doublet model is indeed possible, and
the example can be found, for example, in Ref.~\cite{Koide-Sato}.

The Hermitian matrix $H = M M^{\dagger}$ is diagonalized 
by a unitary matrix $U_{L}$ as
\begin{equation}
U^{\dagger}_{L} H U_{L} = {\rm diag} (m^2_{1}, m^2_{2}, 
m^2_{3})  \ ,
%\eqno(2.5)
\end{equation}
\begin{equation}
U_{L} = P  R  \ ,
%\eqno(2.6)
\end{equation}
\begin{equation}
R =
\left(
\begin{array}{ccc}
c & s & 0   \\
-\frac{1}{\sqrt2} s & \frac{1}{\sqrt2} c & -\frac{1}{\sqrt2} \\
-\frac{1}{\sqrt2} s & \frac{1}{\sqrt2} c & \frac{1}{\sqrt2}
\end{array} \right)  \ ,
%\eqno(2.7)
\end{equation}
\begin{equation}
s = \sin \theta = \sqrt{\frac{m_{1}}{m_{1} + m_{2}}} \ , \ \ 
c = \cos \theta = \sqrt{\frac{m_{2}}{m_{1} + m_{2}}} \ ,
%\eqno(2.8)
\end{equation}
\begin{eqnarray}
-m_{1}& =& \frac{1}{2} \left[b(1+x) - \sqrt{8a^2 + b^2(1+ x)^2}
\,\right] \  , \nonumber \\
m_{2} &=& \frac{1}{2} \left[b(1+x) + \sqrt{8a^2 + b^2(1+ x)^2}
\,\right] \  , \\
%\eqno(2.9)
m_{3} &=& b(1-x) \ , \nonumber
\end{eqnarray}
where $a$, $b$ and $x$ are real parameters given in Eq.~(1.2)
\cite{univtex}. 
When we consider 
$m_{3} >m_{2} > m_{1}$, we can obtain the 
Cabibbo--Kobayashi--Maskawa \cite{CKM} (CKM) matrix  $V$,
\begin{equation}
V = U^{\dagger}_{uL} U_{dL} = R^T_u P R_d =
\left(
\begin{array}{ccc}
c_u c_d + \rho s_u s_d & c_u s_d - \rho s_u c_d & -\sigma s_u \\
s_u c_d - \rho c_u s_d & s_u s_d + \rho c_u c_d & \sigma c_u  \\
-\sigma s_d            & \sigma c_d             & \rho
\end{array} \right) \ ,
%\eqno(2.10)
\end{equation}
where
\begin{equation}
P = P^{\dagger}_u P_d = {\rm diag} (e^{i \delta_1}, e^{i \delta_2},
e^{i \delta_3}) \ ,
%\eqno(2.11)
\end{equation}
\begin{equation}
\rho = \frac{1}{2} (e^{i \delta_3} + e^{i \delta_2} ) = \cos 
\frac{\delta_3 - \delta_2}{2} {\rm exp}\, i 
\left(\frac{\delta_3 + \delta_2}{2} \right) \ ,
%\eqno(2.12)
\end{equation}
\begin{equation}
\sigma = \frac{1}{2} (e^{i \delta_3} - e^{i \delta_2}) = \sin 
\frac{\delta_3 - \delta_2}{2} {\rm exp}\, i 
\left(\frac{\delta_3 + \delta_2}{2} + \frac{\pi}{2} \right) \ ,
%\eqno(2.13)
\end{equation}
where we have taken $\delta_1=0$ without losing generality.
The result (2.10) leads to the following phase-parameter-independent
predictions \cite{Branco}
\begin{equation}
\left| \frac{V_{ub}}{V_{cb}} \right| = \frac{s_u}{c_u} 
= \sqrt{\frac{m_u}{m_c}} = 0.0586 \pm 0.0064  \ ,
%\eqno(2.14)
\end{equation}
\begin{equation}
\left| \frac{V_{td}}{V_{ts}} \right| = \frac{s_d}{c_d} 
= \sqrt{\frac{m_d}{m_s}} = 0.224 \pm 0.014  \ ,
%\eqno(2.15)
\end{equation}
where we have used the values  \cite{F-K} at $\mu=m_Z$ as the quark
mass values.
Although the prediction (2.14) is somewhat small compared with 
the observed 
value \cite{PDG} $|V_{ub} / V_{cd}| = (3.6 \pm 0.7)\times 10^{-3} 
/ (4.12 \pm 2.0)\times 10^{-2}  \simeq 0.087$, 
the prediction (2.10) is satisfactory, roughly speaking. 
For the neutrino mass matrix
$M_\nu$, by taking $\delta_3 - \delta_2 = \pi / 2$, we can obtain 
\cite{univtex} a satisfactory prediction of the lepton mixing matrix 
$V_\ell = U^{\dagger}_e U_\nu$
with a nearly bimaximal mixing.

On the other hand, very recently, 
it has been pointed out by Matsuda and Nishiura \cite{Nishiura} that 
if we assign the up-quark masses as $(m_{u1}, m_{u2}, m_{u3})=
(m_u,m_t,m_c)$ (they have called it Type B) in contrast to the assignment 
$(m_{d1}, m_{d2}, m_{d3})=(m_d, m_s, m_b)$ (Type A) in the mass eigenvalues 
(2.9), then we can obtain phase-parameter-independent relations
\begin{equation}
\left| \frac{V_{ub}}{V_{tb}} \right| = \frac{s_u}{c_u} 
= \sqrt{\frac{m_u}{m_t}} =  (3.6 \pm 0.5) \times 10^{-3}  \ ,
%\eqno(2.16)
\end{equation}
\begin{equation}
\left| \frac{V_{cd}}{V_{cs}} \right| = \frac{s_d}{c_d} 
= \sqrt{\frac{m_d}{m_s}} = 0.224 \pm 0.014  \ ,
%\eqno(2.17)
\end{equation}
instead of the relations (2.14) and (2.15). 
(We will refer this model as Model II.) 
The relation (2.16) is in excellent agreement with the observed value
\cite{PDG} 
$|V_{ub}| = (3.6 \pm 0.7) \times 10^{-3}$, because we have known 
$|V_{tb}| \simeq 1$. The relation (2.17) is consistent with the well-known
relation \cite{Vus} $|V_{us}| \simeq \sqrt{m_d / m_s}$, 
because we have known $|V_{cs}| \simeq 1$ and $|V_{cd}| \simeq |V_{us}|$ .

Thus, the new assignment of the quark masses by Matsuda and Nishiura 
seems to
be favorable phenomenologically. 
However, why is such different assignment 
between the up- and down-quark masses caused?

When such the inverse assignment is caused in the up-quark sector, 
the up-quark mixing matrix $U_{uL}$ is given by
\begin{equation}
U_{uL} = P_u R_u T_{23} \ ,
%\eqno(2.18)
\end{equation}
where
\begin{equation}
T_{23} =
\left(
\begin{array}{ccc}
1 & 0 & 0 \\
0 & 0 & 1 \\
0 & 1 & 0
\end{array} \right)\ ,
%\eqno(2.19)
\end{equation}
so that the CKM mixing matrix $V$ is given by
\begin{equation}
V = T_{23} R^T_u P R_d = 
\left(
\begin{array}{ccc}
c_u c_d + \rho s_u s_d & c_u s_d - \rho s_u c_d & -\sigma s_u \\
-\sigma s_d            & \sigma c_d             & \rho \\
s_u c_d - \rho c_u s_d & s_u s_d + \rho c_u c_d & \sigma c_u  \\
\end{array} \right)\ ,
%\eqno(2.20)
\end{equation}
instead of the relation (2.10).
The new CKM matrix (2.20) predicts 
\begin{equation}
|V_{cb} | = | \rho | = \cos \frac{\delta_3 - \delta_2}{2}\ ,
%\eqno(2.21)
\end{equation}
instead of the old prediction $|V_{cb}| \simeq \sin 
(\delta_3 - \delta_2) / 2 $. In order to give the observed value
$|V_{cb}|=0.0412$, we must take $\delta_3 - \delta_2 = \pi - \varepsilon$
with $\varepsilon = 4.27^\circ$. In the mass matrix form (1.1), 
the phase matrix $P_f$
has been introduced as a measure of the phenomenological 
$2\leftrightarrow 3$ symmetry breaking. (We have assumed that the
$2\leftrightarrow 3$ symmetry is broken only by the phase parameters.)
Therefore, it is natural to consider that such phase parameters 
$\delta^{f}_i$ show $|\delta^{f}_i| \ll 1$. What is the origin of 
such a large value $\delta_3 - \delta_2 \simeq \pi$?

%%%%%%%%%%%%%%%%%%%%%%%%%%%%%%%%%%%%%%%%%%%%%%%%%
\section{Universal texture of quark and lepton mass matrices}

Stimulated by Model II \cite{Nishiura}, in the present section, 
let us speculate a new universal texture of the quark and lepton 
mass matrices.

We consider that the different assignment between the up- and down-quark
masses in Model II is caused by the difference of the initial values of the
parameters $a_f$, $b_f$ and $c_f$ between the up-  and down-quark
sectors in the texture (1.2). 
In fact, the mass hierarchies $m_3 > m_2 > m_1$ or $m_2 > m_3 > m_1$ take
place according as $x_f<0$ or $x_f>0$, respectively, for $b_f \gg a_f > 0$.
Therefore, in order to give the assignment 
$(m_1, m_2, m_3) = (m_u, m_t, m_c)$,
we take the up-quark mass matrix $M_u$ as 
\begin{equation}
\widehat{M}_u = a_u
\left(
\begin{array}{ccc}
0 & 1 & 1 \\
1 & 0 & 0 \\
1 & 0 & 0
\end{array} \right)
 + b_u
\left(
\begin{array}{ccc}
0 & 0       & 0 \\
0 & 1       & 1-\xi_u \\
0 & 1-\xi_u & 1
\end{array} \right)\ ,
%\eqno(3.1)
\end{equation}
where we have put $x_f = 1 - \xi_f$ ($\xi_f$ is a small positive
parameter). 
Similarly, if we want to give $(m_1, m_2, m_3) = (m_d, m_s, m_b)$, 
we should take
\begin{equation}
\widehat{M}_d = a_d
\left(
\begin{array}{ccc}
0 & 1 & 1 \\
1 & 0 & 0 \\
1 & 0 & 0
\end{array} \right)
 + b_d
\left(
\begin{array}{ccc}
0 & 0       & 0 \\
0 & 1       & -(1-\xi_d) \\
0 & -(1-\xi_d) & 1
\end{array} \right) \ \ .
%\eqno(3.2)
\end{equation}
{}From Eqs.~(3.1) and (3.2), the following general form of 
$\widehat{M}_f$ is suggested:
\begin{equation}
\widehat{M}_f = \left(
\begin{array}{ccc}
0  & a_f & a_f \\
a_f  & b_f & (1- \xi_f)b_f e^{i \beta_f} \\
a_f  & (1- \xi_f)b_f e^{i \beta_f} & b_f 
\end{array} \right) \ .
%\eqno(3.3)
\end{equation}

On the other hand, we must consider the origin of $\delta_3
- \delta_2 \simeq \pi$, which is required in Eq.~(2.21) in Model
II. Therefore, we extend the flavor $2 \leftrightarrow
3$ symmetry which is generated by the operator $T_{23}$ , (2.19), 
to a generalized $2 \leftrightarrow 3$ symmetry which is generated by
\begin{equation}
T_{23}^{\phi} = \left(
\begin{array}{ccc}
1 & 0 & 0 \\
0 & 0 & e^{-i \phi} \\
0 & e^{i \phi} & 0 
\end{array} \right) \ ,
%\eqno(3.4)
\end{equation}
as
\begin{equation}
T_{23}^{\phi} M (T_{23}^{\phi})^T = M \ .
%\eqno(3.5)
\end{equation}
(In the present model, too, we assume that the mass matrix $M$ is symmetric,
i.e. $M^T = M$.) The requirement (3.5) leads to relations 
\begin{equation}
M_{12} = M_{13} e^{-i \phi} \ , \ \ M_{22} = M_{33} e^{-2i \phi} \ ,
%\eqno(3.6)
\end{equation}
but the relative phases among $M_{13}$,\ $M_{23}$ and $M_{33}$ are free. 
Therefore, as trial, we assume that $M_{13}$, \ $M_{23}$ and $M_{33}$ 
have the same phases, i.e. we assume the mass matrix form
\begin{equation}
M = \left(
\begin{array}{ccc}
0 & a e^{-i \phi} & a \\
a e^{-i \phi} & b e^{-2i \phi} & (1 - \xi)b \\
a & (1 - \xi)b & b 
\end{array} \right) \ ,
%\eqno(3.7)
\end{equation}
where $a$ and $b$ are positive parameters of the model. 
This mass matrix (3.7) can give a nearly bimaximal mixing
as we show in Appendix.
However, we have still 4 parameters in the mass matrix (3.7).
We would like to seek for a model with a more concise structure.
As we see in Appendix (A.5), in order to give three different
mass eigenvalues, we may consider either model with $\phi=0$ or
$\xi=0$.  However, if we take a model with $\phi=0$, we must
introduce an alternative phase factor in order to explain the
observed $CP$ violation in the quark sectors.

In the present model, we simply assume a texture with
$\xi=0$ in the texture (3.7):
\begin{equation}
M 
= a  \left(
\begin{array}{ccc}
0 &  e^{-i \phi} & 1 \\
e^{-i \phi} & 0 & 0 \\
1 & 0 & 0 
\end{array} \right) +
b  \left(
\begin{array}{ccc}
0 & 0 & 0 \\
0 & e^{-2i \phi} & 1 \\
0 & 1 & 1 
\end{array} \right)
= \left(
\begin{array}{ccc}
0 & a e^{-i \phi} & a \\
a e^{-i \phi} & b e^{-2i \phi} & b \\
a & b & b 
\end{array} \right)
 \ ,
%\eqno(3.8)
\end{equation}
i.e. we have assumed a democratic form except for phases.
It is convenient to rewrite the texture (3.8) as
\begin{equation}
M = P(0, \ -\phi, \ 0) \cdot \widehat{M} \cdot  P(0, \ -\phi, \ 0) \ \ ,
%\eqno(3.9)
\end{equation}
where
\begin{equation}
\widehat{M} = \left(
\begin{array}{ccc}
0 & a  & a \\
a  & b & b e^{i\phi} \\
a & b e^{i\phi} & b 
\end{array} \right) \ ,
%\eqno(3.10)
\end{equation}
\begin{equation}
P(\delta_1, \ \delta_2, \ \delta_3) 
= {\rm diag} (e^{i \delta_1}, \ e^{i \delta_2}, e^{i \delta_3}) \ \ .
%\eqno(3.11)
\end{equation}
The matrix $\widehat{M}$ ( also $M$) has the following eigenvalues:
\begin{eqnarray}
m_1& =& b \left(\sqrt{\cos^2 \frac{\phi}{2} + 2k^2} - \cos \frac{\phi}{2}
\right )
 \  , \nonumber \\
m_2 &=& b \left(\sqrt{\cos^2 \frac{\phi}{2} + 2k^2} + \cos \frac{\phi}{2}
\right )
 \  , \\ 
m_3 &=& 2b \sin \frac{\phi}{2} \ , \nonumber
%\eqno(3.12)
\end{eqnarray}
where $k=a/b$.
Inversely, from Eq.~(3.12), we can evaluate the parameters $a$, $b$ and 
$\phi$ as follows:
\begin{equation}
a = \sqrt{\frac{m_1  m_2}{2}} \ \ ,
%\eqno(3.13)
\end{equation}
\begin{equation}
b = \frac{1}{2} \sqrt{m^2_3 + (m_2 - m_1)^2} \ \ ,
%\eqno(3.14)
\end{equation}
\begin{equation}
\tan \frac{\phi}{2} = \frac{m_3}{m_2 - m_1} \ \ .
%\eqno(3.15)
\end{equation}

The mixing matrix $\widehat{U}$ for the matrix $\widehat{M}$
is given by
\begin{equation}
\widehat{U} = \widehat{R} \cdot P(\frac{\phi}{4}, \frac{\phi}{4}, 
-\frac{\phi}{4}) \ \ ,
%\eqno(3.16)
\end{equation}
\begin{equation}
\widehat{R} = \left(
\begin{array}{ccc}
c e^{-i \phi/2} & s e^{-i \phi/2} & 0 \\
-\frac{1}{\sqrt2}s & \frac{1}{\sqrt2}c & -\frac{1}{\sqrt2} \\
-\frac{1}{\sqrt2}s & \frac{1}{\sqrt2}c & \frac{1}{\sqrt2} 
\end{array} \right) \ ,
%\eqno(3.17)
\end{equation}
where the mixing matrix $\widehat{U}$ has been defined by
\begin{equation}
\widehat{U}^{\dagger} \widehat{M} \widehat{U}^{*} = D \equiv {\rm diag}
(-m_1, \ m_2, \ m_3)  \ .
%\eqno(3.18)
\end{equation}
Therefore, the mixing matrix $U$ for the matrix $M$ is given by
\begin{equation}
U= P(0,-\phi,0)\cdot \widehat{U} =  P(0,-\phi,0) \cdot
\widehat{R} \cdot P(\frac{\phi}{4},  \frac{\phi}{4},  
-\frac{\phi}{4}) \ .
%\eqno(3.19)
\end{equation}

In order to give the phenomenological value 
$\delta_3 - \delta_2 = \pi -\varepsilon$ in Model II, we assume 
\begin{equation}
\phi_u = \varepsilon_u \ , \ \ \ \phi_d =\pi - \varepsilon_d \ ,
%\eqno(3.20)
\end{equation}
which lead to the mass assignments
$(m_{u1}, m_{u2},m_{u3})=(m_u,m_t,m_c)$ and
$(m_{d1}, m_{d2},m_{d3})=(m_d,m_s,m_b)$, respectively.
Then, we obtain
\begin{equation}
U_u = P(0,  -\varepsilon_u,  0) \cdot \widehat{R}_u \cdot 
P(\frac{\varepsilon_u}{4} ,  \frac{\varepsilon_u}{4} ,  
-\frac{\varepsilon_u}{4} ) \ ,
%\eqno(3.21)
\end{equation}
\begin{equation}
U_d = P(0, \pi -\varepsilon_d, 0) \cdot \widehat{R}_d \cdot 
P(\frac{\pi}{4} - \frac{\varepsilon_d}{4} , 
\frac{\pi}{4} - \frac{\varepsilon_d}{4} ,  
-\frac{\pi}{4} + \frac{\varepsilon_d}{4} ) \ ,
%\eqno(3.22)
\end{equation}
so that we obtain the CKM matrix
\begin{equation}
V= T_{23} U_u^\dagger U_d = T_{23} 
P(-\frac{\varepsilon_u}{4}, -\frac{\varepsilon_u}{4} ,  
\frac{\varepsilon_u}{4} )\cdot \widehat{R}_u^\dagger \cdot
P(0, \pi-\varepsilon_d +\varepsilon_u,  0) \cdot \widehat{R}_d 
\cdot P(\frac{\pi}{4} - \frac{\varepsilon_d}{4} , 
\frac{\pi}{4} - \frac{\varepsilon_d}{4} ,  
-\frac{\pi}{4} + \frac{\varepsilon_d}{4} )  \ ,
%\eqno(3.23)
\end{equation}
which essentially gives the same results as the mixing matrix
(2.20) in Model II (but with 
$\delta_2-\delta_3= \pi-\varepsilon_d +\varepsilon_u$)
as far as $|V_{ij}|$ are concerned. 
In addition to those predictions, we can obtain a new prediction
\begin{equation}
|V_{cd}| = \cos \frac{\phi_d - \phi_u}{2} = \sin 
\frac{\varepsilon_d + \varepsilon_u}{2} \ .
%\eqno(3.24)
\end{equation}
Since, from the formula (3.15), we obtain
\begin{equation}
\tan \frac{\varepsilon_u}{2} = \frac{m_c}{m_t - m_u} \simeq 
\frac{m_c}{m_t} \ ,
%\eqno(3.25)
\end{equation}
\begin{equation}
\tan \frac{\varepsilon_d}{2} = \frac{m_s - m_d}{m_b} \simeq 
\frac{m_s}{m_b} \ ,
%\eqno(3.26)
\end{equation}
we can predict
\begin{equation}
|V_{cb}| \simeq \frac{m_s}{m_b} + \frac{m_c}{m_t} = 0.033 \ ,
%\eqno(3.27)
\end{equation}
by using the quark mass values \cite{F-K} at $\mu = m_Z$. 
The value (3.27) is somewhat smaller compared with the observed value 
\cite{PDG}
$|V_{cd}| = 0.0412 \pm 0.0020$, but it is roughly in agreement with the
experimental value.   
For reference, the parameter values of $a$, $b$ and $\varepsilon$ 
which are estimated from the observed quark
masses at $\mu = m_Z$ are listed in Table I.

%%%%%%%%%%%%%%%%%%%%%%%%%%%%%%%%%%%%%%%%%%%%%%%%%%%Table I
\begin{table}
\caption{
Parameter values evaluated from the quark mass values at $\mu = m_Z$.
For reference, in addition to those in the present model with $\xi =0$, 
those in Model II (with $\varepsilon =0$) are listed.
}
 
\begin{tabular}{|c|c|c|c|c|} \hline
 & & $M_u$ & $M_d$ & $M_e$ \\ \hline
 & $(m_1, \ m_2, \ m_3)$ & $(m_u, \ m_t, \ m_c)$ & $(m_d, \ m_s, \ m_b)$ 
& $(m_e, \ m_\mu, \ m_\tau)$ \\ \hline

{\rm Model} & $a/b$ & 0.00507 & 0.00986 & 0.00572 \\
{\rm with}  & $\varepsilon$ & $\varepsilon_u = \phi_u $ 
& $\varepsilon_d = \pi - \phi_d $ 
& $\varepsilon_e = \pi - \phi_e $ \\
$\xi = 0$ &   & $= 0.00748\ (0.429^{\circ})$ & 
$= 0.0591\ (3.39^{\circ})$ & 
$=0.117\ (6.70^{\circ})$ \\
 & $b$ & 90.5 \ {\rm GeV} & 1.50 \ {\rm GeV} & 0.875 \ {\rm GeV} \\ \hline

{\rm Model} & $a/b$ & 0.00506 & 0.00958 & 0.00541 \\
{\rm with} & $\xi$ & 0.00745 & 0.0574 & 0.111 \\
$\varepsilon =0$ & b & 90.2 \ {\rm GeV} & 1.54 \ {\rm GeV} 
& 0.924 \ {\rm GeV} \\ \hline
\end{tabular}
\end{table}
%%%%%%%%%%%%%%%%%%%%%%%%%%%%%%%%%%%%%%%%%%%%%%%%%%%%%

%%%%%%%%%%%%%%%%%%%%%%%%%%%%%%%%%%%%%%%%%%%%%%%%%
\section{Neutrino mass matrix}

Now, let us investigate the lepton sectors under the ansatz (1.4). 
We again consider that the charged lepton mass matrix $M_e$ is 
given by the texture
(1.4) with $\phi_e = \pi-\varepsilon_e$ as well as $M_d$. 
In Model I, the 
phenomenological parameter $\delta_3 - \delta_2$ in the lepton sector was
required as $\delta_3 - \delta_2 = \pi/2$. 
This suggests $\phi_\nu = \pi/2$
in the present model. At present, there is no reason that we should take
$\phi_\nu = \pi/2$. 
However, from the phenomenological point of view,
it is worth investigating the possibility $\phi_\nu = \pi/2$.

When we consider that the neutrino masses are generated 
by the seesaw mechanism
\cite{seesaw}, the neutrino mass matrix $M_\nu$ is given by $M_\nu = M_D 
M^{-1}_R M^T_D$, where $M_D$ is a Dirac neutrino mass matrix and $M_R$ is a
Majorana mass matrix of right--handed neutrino $\nu_{Ri}$. 
Although the origin of the mass generation of $M_R$ is different from that 
$M_u$,\ $M_d$,\ $M_e$ and $M_D$ which are generated by  Higgs scalars of 
SU(2)$_L$ doublet, since the texture (1.4) is based on the properties of 
flavors of $u_{L/R}$,\ $d_{L/R}$,\ $e_{L/R}$ and $\nu_{L/R}$, it is likely 
that the Majorana mass matrix $M_R$ has also the same texture as the Dirac 
mass matrix $M_D$. 
However, note that even if we assume that the mass matrices 
$M_D$ and $M_R$ are given by the texture (1.4), in general, the matrix 
$M_D  M_R^{-1}  M_D^T$ does not take the texture (1.4). 
Only when we consider $\phi_D =\phi_R$, the expression of
$M_D  M_R^{-1}  M_D^T$ becomes a little simpler.
(In order that $M_D  M_R^{-1}  M_D^T$ has the texture (1.4) completely, 
the parameter values have to
satisfy the relations $a_D/a_R=b_D/b_R$ and $\phi_D=\phi_R$.)
In the present paper, we assume only that 
$\phi_D = \phi_R \equiv \phi_\nu$ and we do not assume
$a_D/a_R=b_D/b_R$. 
Then, we obtain
\begin{equation}
M_\nu = M_D \ M_R^{-1} \ M_D^T
= e^{i (\frac{\phi_\nu}{2} - \frac{\pi}{2} + \varepsilon)} 
P(-\frac{1}{2}\phi_\nu + \frac{1}{2}\pi - \varepsilon \ , -\phi_\nu \ , 0) 
 \cdot
\widehat{M} \cdot 
P(-\frac{1}{2}\phi_\nu + \frac{1}{2}\pi - \varepsilon \ , -\phi_\nu \ , 0) \ ,
%\eqno(4.1)
\end{equation}
\begin{equation}
\widehat{M} = \left(
\begin{array}{ccc}
0 & a & a \\
a & b & be^{i \widehat{\phi}}  \\
a & be^{i \widehat{\phi}} & b 
\end{array} \right) \ ,
%\eqno(4.2)
\end{equation}
\begin{equation}
a \equiv \frac{a^2_D}{a_R} \ ,
%\eqno(4.3)
\end{equation}
\begin{equation}
b \equiv \frac{b^2_D}{b_R} \sin \frac{\phi_\nu}{2} \sqrt{1+r^2 {\rm cot}^2 
\frac{\phi_\nu}{2}}
  \ ,
%\eqno(4.4)
\end{equation}
\begin{equation}
r \equiv 2\frac{a_D b_R}{b_D a_R} - \left(\frac{a_D b_R}{b_D a_R} 
\right)^2  \ ,
%\eqno(4.5)
\end{equation}
\begin{equation}
\tan \varepsilon = r {\rm cot} \frac{\phi_\nu}{2} \ ,
%\eqno(4.6)
\end{equation}
\begin{equation}
\widehat{\phi} = \pi - 2\varepsilon  \ .
%\eqno(4.7)
\end{equation}

Now, we assume $\phi_\nu = \pi /2$ and $a_D / b_D \ll a_R / b_R$, 
(i.e. $r \ll 1$),
so that we obtain
\begin{equation}
\tan \varepsilon = r \ .
%\eqno(4.8)
\end{equation}
Since the mixing matrices $U_e$ and $U_\nu$ are given by
\begin{equation}
U_{eL} = P(0, \ -\phi_e, \ 0) \cdot \widehat{R}_e \cdot 
P(\frac{1}{4}\phi_e, \ \frac{1}{4}\phi_e, \ -\frac{1}{4}\phi_e)
\ ,
%\eqno(4.9)
\end{equation}
\begin{equation}
U_{\nu L} = e^{i \frac{1}{2} (\varepsilon - \frac{\pi}{4})}
P(\frac{1}{4}\pi-\varepsilon , \ -\frac{\pi}{2}, \ 0) \cdot 
\widehat{R}_{\nu} \cdot 
P(\frac{1}{4}\widehat{\phi}, \ \frac{1}{4}\widehat{\phi}, \ 
-\frac{1}{4}\widehat{\phi}) \ ,
%\eqno(4.10)
\end{equation}
where $\phi_e = \pi - \varepsilon_e$, \ $\widehat{\phi} 
= \pi - 2\varepsilon$,
and $\widehat{R}_e$ and $\widehat{R}_\nu$ are given by Eq.(3.17),
the lepton mixing matrix $V_{\ell} = U^{\dagger}_{eL} U_{\nu L}$ 
is expressed as follows:
\begin{equation}
V_\ell = \left(
\begin{array}{ccc}
c_e c_\nu e^{i \delta_1} + \rho s_e s_\nu & 
c_e s_\nu e^{i \delta_1} - \rho s_e c_\nu & 
-\sigma s_e \\
s_e c_\nu e^{i \delta_1} - \rho c_e s_\nu & 
s_e s_\nu e^{i \delta_1} + \rho c_e c_\nu &
\sigma c_e \\
- \sigma s_\nu & 
\sigma c_\nu & 
\rho 
\end{array} \right) \ ,
%\eqno(4.11)
\end{equation}
where
\begin{equation}
s_e = \sqrt{\frac{m_e}{m_\mu + m_e}} \ , \ \ \ 
c_e = \sqrt{\frac{m_\mu}{m_\mu + m_e}} \ , 
%\eqno(4.12)
\end{equation}
\begin{equation}
s_\nu = \sqrt{\frac{m_{\nu 1}}{m_{\nu 2} + m_{\nu 1}}} \ , \ \ \ 
c_\nu = \sqrt{\frac{m_{\nu 2}}{m_{\nu 2} + m_{\nu 1}}} \ , 
%\eqno(4.13)
\end{equation}
\begin{equation}
\delta_1 = \frac{\pi}{4} - \varepsilon + \frac{\widehat{\phi}}{2}
- \frac{\phi_e}{2}  \ ,
%\eqno(4.14)
\end{equation}
and $\rho$ and $\sigma$ are defined by Eqs.~(2.12) and (2.13) with
$\delta_2 = \phi_e - \pi = \varepsilon_e - \pi/2$ and $\delta_3 =0$. 
Exactly
speaking, the mixing matrix $V_{\ell}$ is given by
\begin{equation}
V_\ell=e^{i \frac{1}{2} (\varepsilon - \frac{\pi}{4})}
P(-\frac{1}{4}\phi_e , \ -\frac{1}{4}\phi_e , \ 
\frac{1}{4}\phi_e) \cdot \widehat{V}_{\ell} \cdot 
P(\frac{1}{4}\widehat{\phi}, \ \frac{1}{4}\widehat{\phi}, \ 
-\frac{1}{4}\widehat{\phi}) \ ,
%\eqno(4.15)
\end{equation}
where $\widehat{V_\ell}$ in the expression (4.15) is defined by $V_\ell$
in Eq.~(4.11). As far as the magnitudes of $(V_\ell)_{ij}$ are concerned, 
we can drop the phase factors 
$e^{i \frac{1}{2} (\varepsilon - \frac{\pi}{4})}
P(-\frac{1}{4}\phi_e ,  -\frac{1}{4}\phi_e , \frac{1}{4}\phi_e)$ and 
$P(\frac{1}{4}\widehat{\phi},  \frac{1}{4}\widehat{\phi}, 
-\frac{1}{4}\widehat{\phi})$. Of course, when we deal with 
neutrinoless double beta decay, a $CP$ violation process, and so on, 
we must exactly take those phase factors into consideration. 
For a time, since we discuss 
$\sin^2 2\theta_{atm}$ and $\tan^2 \theta_{solar}$, 
we neglect those phase factors.

For $\sin^2 2\theta_{atm}$, we obtain
\begin{eqnarray}
\sin^2 2\theta_{atm} & = & 4|(V_\ell)_{23}|^2 |(V_\ell)_{33}|^2 = 4|\sigma|^2 
|\rho|^2 c^2_e = \frac{m_\mu}{m_\mu + m_e} 
\sin^2(\frac{\pi}{2} - \varepsilon_e) \nonumber \\
 & \simeq & 1 - \frac{m_e}{m_\mu} - 4\left(\frac{m_\mu}{m_\tau}\right)^2 
= 0.98 \ \ .
%\eqno(4.16)
\end{eqnarray}
We also obtain
\begin{equation}
|(V_\ell)_{13}|^2 = |\sigma|^2 s^2_e  \simeq \frac{m_e}{2 m_\mu} 
\left(1- 2\frac{m_\mu}{m_\tau} \right) = 0.0021 \ \ .
%\eqno(4.17)
\end{equation}
These values (4.16) and (4.17) are consistent with the observed values 
\cite{atm,CHOOZ}.

For $\tan^2 \theta_{solar}$, we obtain as follows. 
Note that in the expression
$(V_\ell)_{ij} \ \ (i,j = 1,2)$ given by Eq.~(4.11), 
the relative phase of the
first term to the second term, 
$\delta_1 - (\delta_3 + \delta_2) /2$, is given by
\begin{equation}
\delta_1 - \frac{\delta_3 + \delta_2}{2} = \left(\frac{\pi}{4}-\varepsilon 
-\frac{1}{2}\widehat{\phi} + \frac{1}{2}\phi_e\right) - 
\frac{1}{2} \left(-\frac{\pi}{2} + \phi_e\right) =0 \ \ ,
%\eqno(4.18)
\end{equation}
so that we can write $(V_\ell)_{ij}$ as 
\begin{equation}
(V_{\ell})_{11} = (c_e c_\nu + |\rho| s_e s_\nu) e^{i \delta_1}  \ \ ,
%\eqno(4.19)
\end{equation}
\begin{equation}
(V_{\ell})_{12} = (c_e s_\nu - |\rho| s_e c_\nu) e^{i \delta_1}  \ \ ,
%\eqno(4.20)
\end{equation}
and so on. Therefore, we obtain
\begin{equation}
\tan^2 \theta_{solar} = \frac{|(V_{\ell})_{12}|^2}{|(V_{\ell})_{11}|^2}
= 
\left(\frac{{\frac{s_\nu}{c_\nu} - |\rho| \frac{s_e}{c_e}}}
{1+|\rho| \frac{s_e}{c_e} \frac{s_\nu}{c_\nu}} \right)^2 \simeq 
\left(\frac{s_\nu}{c_\nu} \right)^2 = \frac{m_{\nu 1}} {m_{\nu 2}}
\ \ ,
%\eqno(4.21)
\end{equation}
for $s_\nu / c_\nu = \sqrt{m_{\nu 1} / m_{\nu 2}} \gg s_e / c_e = 
\sqrt{m_e / m_\mu}$.
(The alternative case $s_\nu / c_\nu \leq s_e / c_e$ is ruled out 
because the case leads to a very small value of $\tan^2
\theta_{solar}$.)

For $\Delta m^2_{solar}$ and $\Delta m^2_{atm}$, from Eq.~(3.12) with
$\phi = \widehat{\phi} = \pi -2\varepsilon$, we can obtain
\begin{equation}
\Delta m^2_{21} \equiv m^2_2 - m^2_1 = 4b^2 \sin \varepsilon
\sqrt{\sin^2 \varepsilon + 2k^2}  \ \ ,
%\eqno(4.22)
\end{equation}
\begin{equation}
\Delta m^2_{32} \equiv m^2_3 - m^2_2 = 4b^2 \left(1-\frac{3}{2} \sin^2
\varepsilon - \frac{1}{2} k^2 -2\sin \varepsilon
\sqrt{\sin^2 \varepsilon + 2k^2} \right)  \ \ ,
%\eqno(4.23)
\end{equation}
where
\begin{equation}
k \equiv \frac{a}{b} = \sqrt2 \frac{a^2_D b_R}{b^2_R a_R} 
\cos \varepsilon \ \ ,
%\eqno(4.24)
\end{equation}
so that we predict
\begin{equation}
R \equiv \frac{\Delta m^2_{21}}{\Delta m^2_{32}} \simeq \sin \varepsilon
\sqrt{\sin^2 \varepsilon + 2k^2} \ \ ,
%\eqno(4.25)
\end{equation}
for small $\varepsilon^2$ and $k^2$. 
On the other hand, from 
Eqs.~(3.13) -- (3.14)  [i.e. 
$\tan\varepsilon =(m_{\nu 2}-m_{\nu 1})/m_{\nu 3}$], we obtain
\begin{equation}
k \equiv \frac{a}{b} = \sqrt{\frac{2m_{\nu 1} m_{\nu 2}}
{m^2_{\nu 3} + (m_{\nu 2} - m_{\nu 1})^2 } } = \sqrt{2}
\frac{\sqrt{m_{\nu 1} / m_{\nu 2}}}{1-m_{\nu 1} / m_{\nu 2}} 
\sin \varepsilon
\ ,
%\eqno(4.26)
\end{equation}
Therefore, if we give a value
\begin{equation}
x \equiv  \frac{k}{\sin \varepsilon} \ ,
%\eqno(4.27)
\end{equation}
we can obtain the value of $m_{\nu 1} /m_{\nu 2}$ as follows:
\begin{equation}
\tan^2 \theta_{solar} \simeq \frac{m_{\nu 1}}{m_{\nu 2}} = \frac{1}{x^2}
\left(1+x^2 - \sqrt{1+2x^2} \right) \ .
%\eqno(4.28)
\end{equation}

In the present model, since the value of $R \equiv \Delta m^2_{21} / 
\Delta m^2_{32}$ depends on the parameters $\varepsilon$ and $x$, 
the value of $\tan^2\theta_{solar}$ cannot be predicted from 
the charged lepton masses only. 
In other words, if we give 
the values $R$ and $\tan^2
\theta_{solar}$, we can determine the values $x$ and 
$\varepsilon$ ($k$ and 
$\varepsilon$), so that we can also determine the value of $m_{\nu 1}$, 
$m_{\nu 2}$ and $m_{\nu 3}$. 
In Table II, we list the numerical predictions
$m_{\nu 1}/m_{\nu 2} = \tan^2 \theta_{solar}$ and 
$(m_{\nu 1}, \ m_{\nu 2}, \ m_{\nu 3})$ for typical values of $x= k/\sin
\varepsilon$, where we have used the input values \cite{atm,solar,kamland}
\begin{equation}
R_{obs} = \frac{6.9 \times 10^{-5} \ \ {\rm eV^2}}
{2.5 \times 10^{-3} \ \ {\rm eV^2}} = 2.76 \times 10^{-2} \ ,
%\eqno(4.29)
\end{equation}
and $\Delta m^2_{atm} = 2.5 \times 10^{-3} \ \ {\rm eV^2}$. 
The value $m_{\nu 1}/m_{\nu 2}$ in Table II has been evaluated 
from Eq.~(4.26).
{}From the relation
\begin{equation}
R = \left(\frac{m_{\nu 2}}{m_{\nu 3}} \right)^2 
\frac{1-(\tan^2 \theta_{solar})^2}{1-(m_{\nu 2}/m_{\nu 3})^2} \ ,
%\eqno(4.30)
\end{equation}
we obtain
\begin{equation}
r_{23} \equiv \frac{m_{\nu 2}}{m_{\nu 3}} 
= \sqrt{\frac{R}{1-\tan^4 \theta_{solar} + R}} \ .
%\eqno(4.31)
\end{equation}
The mass values $m_{\nu i}$ in Table II have been obtained from
\begin{equation}
m_{\nu 3} = \sqrt{\frac{\Delta m^2_{atm}}{1-r^2_{23}}} \ ,
%\eqno(4.32)
\end{equation}
$m_{\nu 2} = r_{23} m_{\nu 3}$ and 
$m_{\nu 1} = m_{\nu 2} \tan^2 \theta_{solar}$. 
The value of $\tan \varepsilon$
in Table II has been estimated, not from the approximate 
relation (4.25), but 
from the exact relation
\begin{equation}
\tan \varepsilon = \frac{m_{\nu 2}}{m_{\nu 3}} 
\left(1- \frac{m_{\nu 1}}{m_{\nu 2}} \right) 
= r_{23} (1- \tan^2 \theta_{solar}) \ .
%\eqno(4.33)
\end{equation}

%%%%%%%%%%%%%%%%%%%%%%%%%%%%%%%%%%%%%%%%%%%%%%%%%%%%% Table II
\begin{table}
\caption{
Values $\tan^2 \theta_{solar}$ and $m_{\nu i}$ ($i=1,2,3$)
 for typical value of $x \equiv k /\sin \varepsilon$
}

\begin{tabular}{|c|c|c|c|c|} \hline
$x$ & $m_{\nu 1}/m_{\nu 2} \simeq \tan^2 \theta_{solar}$ 
& $r_{23} =  {m_{\nu 2}}/{m_{\nu 3}}$ 
& $m_{\nu 1} \ {\rm [eV]}$ \ \ \ $m_{\nu 2} \ {\rm [eV]}$ 
\ \ \ $m_{\nu 3} \ {\rm [eV]}$ & $\tan \varepsilon$
\\ \hline
2 & 1/2=0.5 & 0.188 & 0.0048 \ \ \ 0.0096 \ \ \ 0.0509 & 0.094 \\
$\sqrt2$ & $(3-\sqrt5)/2=0.382$ & 0.177 & 0.0034 \ \ \ 0.0090 \ \ \ 0.0508 
& 0.109 \\
$\sqrt{3/2}$ & $1/3=0.333$ & 0.174 & 0.0029 \ \ \ 0.0088 \ \ \ 0.0508 & 0.116 
\\ \hline
\end{tabular}
\end{table} 

%%%%%%%%%%%%%%%%%%%%%%%%%%%%%%%%%%%%%%%%%%%%%%%%%%%%%%%

%%%%%%%%%%%%%%%%%%%%%%%%%%%%%%%%%%%%%%%%%%%%%%%%%
\section{Summary}

In conclusion, we have proposed a universal texture (1.4) 
of quark and lepton mass matrices. 
The mass matrix $M$ is invariant under the extended flavor 
$2 \leftrightarrow 3$ permutation $T^{\phi}_{23}$, (3.4), 
as $T^{\phi}_{23} M
(T^{\phi}_{23})^T =M$. Besides, the matrix elements 
$M_{ij} \ (i =2,3)$ are 
exactly democratic apart from their phases, i.e. 
\begin{equation}
M_f = a_f
\left(
\begin{array}{ccc}
0 & e^{-i \phi_f} & 1 \\
e^{-i \phi_f} & 0 & 0 \\
1 & 0 & 0
\end{array} \right)
\langle H^0_A \rangle + b_f
\left(
\begin{array}{ccc}
0 & 0 & 0 \\
0 & e^{-2i \phi_f} & 1 \\
0 & 1 & 1
\end{array} \right)
\langle H^0_B \rangle\ .
%\eqno(5.1)
\end{equation}
The mass matrix $M$ is described by two parameters $\phi$ and $a/b$, 
as for as 
the mass ratios and mixings are concerned.

For quark sectors $M_u$ and $M_d$, we take the parameter $\phi$ 
as $\phi_u =
\varepsilon_u$ and $\phi_d = \pi -\varepsilon_d$, respectively, where 
$\varepsilon_u$ and $\varepsilon_d$ are small positive parameters. 
Then, we can obtain successful relations for the CKM mixing parameters 
in terms of quark masses. 
(Note that, in Models I and II, the value of $|V_{cb}|$ is given by a 
phenomenological parameter $(\delta_3 - \delta_2)$ independently of 
the quark 
mass ratios, while, in the present model, $|V_{cb}|$ is 
given in terms of quark
mass rations as shown in Eq.~(3.27).)
For the charged lepton mass matrix $M_e$, we take $\phi_e = \pi -
\varepsilon_e$ as well as $\phi_d =\pi -\varepsilon_d$, 
while, for the neutrino
mass matrix $M_\nu$, we take $\phi_\nu = \pi/2$ in order to give a nearly 
bimaximal mixing. The neutrino mass matrix $M_\nu$ is described by two
parameters $\varepsilon$ and $k \equiv a/b$. The predictions 
$\sin^2 2\theta_{atm}$ and $|(V_{\ell})_{13}|^2$ are given only 
in terms of 
charged lepton masses independently of the parameters 
$\varepsilon$ and $k$,
as shown in Eqs.~(4.16) and (4.17). 
These predictions are favorable to the 
data. On the other hand, the quantities $\tan^2 \theta_{solar}$, 
$R \equiv 
\Delta m^2_{solar} / \Delta m^2_{atm}$ and $m_{\nu i}$ ($i=1,2,3$) 
are dependent
on the parameters $\varepsilon$ and $k$ in the neutrino mass matrix 
$M_\nu$. 
In Table II, we have listed the predictions for typical values of 
$x \equiv k/ \sin \varepsilon$.

Although we have taken $\phi_e = \pi - \varepsilon_e$ for 
the charged lepton 
sector, it is not essential. If we take $\phi_e = \varepsilon_e$ 
as well as 
$\phi_u$, the results for the neutrino mixing are substantially 
uncharged (e.g.
Eqs.~(4.16) and (4.17) become merely
$\sin^2 2\theta_{atm} \simeq 1- m_e/m_\tau$ and
$|(V_{\ell})_{13}|^2 \simeq m_e/2m_\tau$, respectively). 
However, the choice 
$\phi_\nu = \pi/2$ is essential. If we choose another value of
 $\phi_\nu$, we
cannot obtain $\sin^2 2\theta_{atm} \simeq 1$. 
It is an open question why we 
must choose $\phi= \pi/2$ only for the neutrino sector.

Since as seen in Table I the parameter values of $\phi$ (and $a/b$) 
which are
determined from the observed fermion masses are different 
from each other (i.e. $a_e \neq d_d$, $b_e \neq b_d$ and
$\phi_e \neq \phi_d$), the present model cannot be applied 
to a grand unification theory (GUT) model straightforwardly. 
However, for example, in an SU(5) GUT model, with matter 
fields $\overline{5} + 10 + \overline{5}' + 5'$ \cite{5-bar5}, 
we can consider a 
$\overline{5} \leftrightarrow \overline{5}'$ mixing. 
Therefore, the problem that the
parameter values are not universal is not so serious defect 
even for a GUT model. 
Because of its simpleness of the texture (1.4) with few parameters, 
it will be worthwhile 
taking the present universal texture seriously, 

%%%%%%%%%%%%%%%%%%%%%%%%%%%%%%%%%%%%%%%%%%%%%%

%\newpage
\vspace{15mm}

%\nonum
\centerline{\large\bf Appendix}

It is convenient to rewrite the mass matrix 
%\begin{equation}
$$
M = \left(
\begin{array}{ccc}
0 & a e^{-i \phi} & a \\
a e^{-i \phi} & b e^{-2i \phi} & (1 - \xi)b \\
a & (1 - \xi)b & b 
\end{array} \right) \ , 
\eqno(A.1)
$$
%\end{equation}
as 
%\begin{equation}
$$
M = P(0,  -\phi,  0) \cdot \widehat{M} \cdot  P(0,  -\phi,  0)  \ ,
\eqno(A.2)
$$
%\end{equation}
where
%\begin{equation}
$$
P(\delta_1,  \delta_2,  \delta_3) 
= {\rm diag} (e^{i \delta_1},  e^{i \delta_2}, e^{i \delta_3})  \ ,
\eqno(A.3)
$$
%\end{equation}
%\begin{equation}
$$
\widehat{M} = \left(
\begin{array}{ccc}
0 & a  & a \\
a  & b  & (1 - \xi)b e^{i \phi} \\
a & (1 - \xi)b e^{i \phi} & b 
\end{array} \right) \ . 
\eqno(A.4)
$$
%\end{equation}

The eigenvalues $m_i$ and mixing matrix $\widehat{U}$ of the 
mass matrix $\widehat{M}$ are follows:
%\begin{eqnarray}
$$
\begin{array}{ll}
 & m_1 =  \frac{1}{2} (\sqrt{p^2 + 8k^2} - p)b  \ , \nonumber \\
 & m_2  =  \frac{1}{2} (\sqrt{p^2 + 8k^2} + p)b  \ ,  \\
 & m_3  =  qb  \ , \nonumber 
\end{array}
\eqno(A.5)
$$
%\end{eqnarray}
%\begin{equation}
$$
\widehat{U} = \widehat{R} \cdot P(-\frac{\delta}{2},  -\frac{\delta}{2},  
\frac{\delta}{2})  \ ,
\eqno(A.6)
$$
%\end{equation}
%\begin{equation}
$$
\widehat{R} = \left(
\begin{array}{ccc}
c e^{i \delta} & s e^{i \delta} & 0 \\
-\frac{1}{\sqrt2}s & \frac{1}{\sqrt2}c & -\frac{1}{\sqrt2} \\
-\frac{1}{\sqrt2}s & \frac{1}{\sqrt2}c & \frac{1}{\sqrt2} 
\end{array} \right) \ ,
\eqno(A.7)
$$
%\end{equation}
where $k=a/b$,
%\begin{equation}
$$
p^2 = \xi^2 + 4(1- \xi) \cos^2 \frac{\phi}{2}  \ ,
\eqno(A.8)
$$
%\end{equation}
%\begin{equation}
$$
q^2 = \xi^2 + 4(1- \xi) \sin^2 \frac{\phi}{2}  \ ,
\eqno(A.9)
$$
%\end{equation}
%\begin{equation}
$$
\tan \delta = -\frac{(1- \xi) \sin \phi}{1+ (1- \xi) \cos \phi}  \ ,
\eqno(A.10)
$$
%\end{equation}
and the mixing matrix $\widehat{U}$ is defined by
%\begin{equation}
$$
\widehat{U}^{\dagger} \widehat{M} \widehat{U}^{*} = D \equiv {\rm diag}
(-m_1, m_2, m_3) \ \ .
\eqno(A.11)
$$
%\end{equation}
Therefore, the mixing matrix $U$ for the matrix $M$ is given by
%\begin{equation}
$$
U= P(0,-\phi,0)\cdot \widehat{U} =  P(0,-\phi,0) \cdot
\widehat{R} \cdot P(-\frac{\delta}{2}, - \frac{\delta}{2},  
\frac{\delta}{2}) \ .
\eqno(A.12)
$$
%\end{equation}

If we put $\phi=0$, then the results (A.5) -- (A.10) become those
in Models I and II.
The model (1.4) in the present paper corresponds to one with $\xi=0$.

%%%%%%%%%%%%%%%%%%%%%%%%%%%%%%%%%%%%%%%%%%%%

\end{document}